\documentclass[sigconf,nonacm]{acmart}
\renewcommand\footnotetextcopyrightpermission[1]{}

\usepackage{xspace}
\usepackage{xcolor}
\usepackage{listings}
\usepackage[framemethod=tikz]{mdframed}
\usepackage{algorithm}
\usepackage[noend]{algpseudocode}
\usepackage[textsize=tiny,textwidth=0.6in,disable]{todonotes}
\usepackage{enumitem}
\definecolor{codeblue}{rgb}{0,0,1}
\definecolor{red}{rgb}{0.6,0,0}
\definecolor{blue}{rgb}{0,0,0.6}
\definecolor{green}{rgb}{0,0.8,0}
\definecolor{lightorange}{RGB}{255,247,230}
\definecolor{BlackColor}{RGB}{0,0,0}
\definecolor{CommentColor}{rgb}{0.12,0.38,0.18}

\lstdefinestyle{P4}{
  showspaces=false,
  showtabs=false,
  tabsize=2,
  columns=flexible,
  keepspaces=true,
  language={Java},
  numbers=left,
  basicstyle=\ttfamily\footnotesize,
  commentstyle=\color{CommentColor}\ttfamily\footnotesize,
  stringstyle=\color{CommentColor},
  escapeinside={/*@}{@*/},
  numberstyle=\scriptsize\color{gray},
  showstringspaces=false,
  upquote=true,
  xleftmargin=1.2em,
  framexleftmargin=1.5em,
  keywords={ StateMachine },
  keywords=[2]{ bool, str, SM },
  keywords=[3]{ assume, guarantee},
  keywords=[4]{ read, assert, call, write},
  keywordstyle=\color{BlackColor}\bfseries,
  keywordstyle=[2]\color{codeblue},
  keywordstyle=[3]\color{red},
  keywordstyle=[4]\color{green},
  moredelim=[il][\color{darkgray}]{$$},
}
\mdfdefinestyle{background}{backgroundcolor=lightorange,innerrightmargin=0cm,innertopmargin=-0.1cm,innerbottommargin=-0.10cm,leftmargin=+0cm, roundcorner=2pt,nobreak=true}

\newcounter{defn}[section]
\renewcommand{\thedefn}{\arabic{section}.\arabic{defn}}
\newenvironment{defn}[2][]{%
\refstepcounter{defn}%
\ifstrempty{#1}%
{\mdfsetup{%
frametitle={%
\tikz[baseline=(current bounding box.east),outer sep=0pt]
\node[anchor=east,rectangle,fill=purple!20]
{\strut Definition~\thedefn};}}
}%
{\mdfsetup{%
frametitle={%
\tikz[baseline=(current bounding box.east),outer sep=0pt]
\node[anchor=east,rectangle,fill=purple!20]
{\strut #1};}}%
}%
\mdfsetup{nobreak=true,innertopmargin=10pt,linecolor=purple!20,%
linewidth=2pt,topline=true,%
frametitleaboveskip=\dimexpr-\ht\strutbox\relax
}
\begin{mdframed}[]\relax%
\label{#2}}{\end{mdframed}}

\newcommand{\sysname}{\textsc{SolVRT}\xspace}

\begin{document}

\title{Synthesizing Voltage Ride-Through Controllers for Data Centers}

\author{%
  Wayne Wang \quad
  Archit Bhatnagar \quad
  Tongyuan Miao \quad
  Saniya Kalamkar \\[4pt]
  Wenqi Cui\textsuperscript{\dag} \quad
  Inigo Incer \quad
  Ang Chen \\[4pt]}

\affiliation{%
  \institution{%
    \textit{University of Michigan} \qquad
    \textsuperscript{\dag}\textit{New York University}
  }
  \city{}\country{}
}

\renewcommand{\shortauthors}{Wang et al.}

\begin{abstract}
Data centers are among the power grid's fastest-growing loads.
Since data center servers are sensitive electronic components, they need to be protected against the grid's voltage disturbances during grid faults.
While this can be achieved by simply disconnecting from the grid upon disturbances, disconnection can further destabilize the power system
if many data centers trip at once. To address this emerging concern,
voltage ride-through (VRT) grid codes have been proposed to standardize data center behavior. They require a data center to stay connected for a period of time through the disturbance, hold an active power floor, and recover its draw within a deadline upon restoration. 
However, systematically designing and certifying controllers that satisfy these coupled temporal and operational requirements remains challenging.

We propose \sysname, a system that synthesizes a grid-code-compliant VRT controller for a given data center using formal methods. 
We develop a specification language that expresses a grid code in Signal Temporal Logic (STL) as the basis for formal reasoning. Our encoding algorithm takes the specification, along with a model of the data center's power topology, and translates the constraints into a controller synthesis problem. This step produces a correct-by-construction controller if a solution can be found, or a proof that no such controller exists. For the latter case, \sysname provides a diagnostic step: it traces the facility's ``conflict frontier,'' isolates the conflicting clauses that led to non-compliance, and computes the smallest hardware or workload change that would enable compliance.
We evaluate \sysname through closed-loop simulations of a 200\,MW data center connected to a 140-bus transmission system. The results demonstrate that \sysname can synthesize compliant VRT controllers, certify infeasibility when compliance is unattainable, and identify targeted modifications that enable compliance.
\end{abstract}

\maketitle

\section{Introduction}

Data centers (DCs) are large electric loads with fluctuating demands and high sensitivity to power quality; their response to grid voltage disturbances has been a rising concern. For instance, if
the grid experiences voltage sags upon contingencies and faults, data centers may simply disconnect from the grid in order to protect internal devices (e.g., by switching to local power sources such as UPSes and diesel generators).
From the grid's perspective, however, simultaneous disconnection of large loads introduces further disturbances during an already-stressed event. 
In a 2024 incident in Virginia, a single transmission-line fault caused voltage sags that tripped roughly 1,500\,MW of data center load at once, most of which stayed offline for hours~\cite{virginia}.\footnote{A larger event of the same type occurred in Virginia on July 22, 2026, with a reported total load loss roughly 3\,GW. Detailed analysis of this incident was not available at the time of writing.}
This has prompted the development of voltage ride-through (VRT) requirements from grid operators, such as ERCOT (US) and Energinet (Europe).
VRT requires the data center to remain connected and follow specified response criteria if the grid voltage sags or swells during a transient fault.

While VRT has been well studied for power devices and generators, data centers present a new range of challenges. First, DCs represent a large \textit{load}, whereas VRT standards were originally designed for \textit{generators}. 
Further, DCs comprise a \textit{system of devices}, each with its own protective response mechanisms invisible to the grid, requiring system-level joint control.
The growth of \textit{AI workloads} compounds these challenges: a single data center campus now draws hundreds of megawatts, its training workload swings within seconds, and the hardware at risk  (e.g., GPU clusters) is costly enough that individual protections are tuned to trip early.

Today, a data center's response to a voltage event heavily depends on individual device protection settings, power electronics vendor behavior, and controller heuristics. 
Under this uncertainty, DC operators default to early disconnection to protect their internal operation. 
A hand-tuned controller might help the DC ride through, but provides no compliance guarantee since grid codes impose several timed requirements that the data center must satisfy simultaneously.
A principled design for the data center to control internal devices and ride through voltage events with guaranteed grid code compliance is still missing. This difficulty is compounded by the fact that the grid codes exist as natural language documents describing numeric constraints, which are hard to analyze formally in a machine-checkable manner. This interface between grid requirements and data center capabilities is described in informal text, making it difficult for  
grid operators to anticipate data center behavior, for data centers to guarantee compliance, or to reason about the hardware or workload provisioning that would achieve compliance under all conditions.

In this paper, we take a fresh look at VRT from the perspective of formal methods. If we can formalize grid codes in a machine-checkable language, a range of synthesis and verification methods becomes available to attack this problem. 
We argue for treating VRT requirements as a formal assume-guarantee specification,
a method widely used in safety-critical domains such as aerospace engineering~\cite{pacti-space,Incer:EECS-2022-99}.
With the assume-guarantee interface, the grid states assumptions on the voltage disturbance, and the data center must uphold guarantees on its response over explicit signals whenever those assumptions hold.
Both assumptions and guarantees are timed constraints on real-valued signals, and Signal Temporal Logic (STL) expresses exactly such predicates over signal trajectories~\cite{maler2004monitoring}.
Compliance can be established by solving whether a controller exists that enables the data center to satisfy the guarantee under the assumption while respecting all constraints.

Following this intuition, we propose a grid-code-aware system \sysname that solves the VRT compliance problem for data centers using formal methods by synthesizing a code-compliant controller for a given data center. 
When this synthesis is infeasible, \sysname diagnoses the unsatisfiable clauses and provides change recommendations.
We make three contributions.   
First, we develop a new domain-specific language (DSL) for expressing data center VRT grid codes. The specification makes voltage event assumptions, response requirements, timers, active-power floor, and internal protection limits explicit and machine-checkable---using STL predicates to capture the core requirements. 
Next, our algorithm encodes the specification into a controller synthesis problem with other constraints, including disturbance traces, plant models, and hardware constraints. When the synthesis query is feasible, \sysname outputs a model predictive controller (MPC) as witness. If synthesis is infeasible, the diagnosis and provisioning step takes over.
We diagnose the violating clauses and provide provisioning recommendations by modeling infeasibility as the \textit{conflict frontier}. This allows \sysname to identify the unsatisfiable core of the conflict and compute the minimal changes needed on hardware capability or workload flexibility to restore compliance for a given data center and specification.

With a comprehensive simulation using a 200MW data center 
and a 140-bus transmission system, we are 
the first to formalize 
VRT requirements for data centers as a controller synthesis problem and demonstrate its practicality. 
We hope this paper will inspire future work that studies DC/grid coordination through the lens of formal analysis to increase the resilience of cross-infrastructure interactions. We will release \sysname in open source.

\section{Background and Related Work}
\label{sec:background}

\subsection{What are VRT and LVRT}

Voltage ride-through (VRT) has traditionally been defined as a generator-side requirement~\cite{tsili2009review,ieee2800}. Its central objective is to ensure that generators, including synchronous machines and inverter-based resources such as solar photovoltaic and wind generation, remain connected during short-duration voltage disturbances caused by grid-side faults. By preventing premature disconnection, VRT requirements help avoid cascading loss of generation and support system stability during and after fault events.

Comparable ride-through requirements have not been imposed on loads, which were typically smaller, more dispersed, and had a less explicitly regulated dynamic voltage response. However, the rapid growth of large electronic loads, particularly DCs, is changing this assumption. Since a single DC today can represent a substantial amount of load, its disconnection during a voltage disturbance can create an abrupt loss of demand and potentially worsen grid dynamics. As a result, large-load ride-through capability is emerging as an important reliability requirement. For example, ERCOT~\cite{ercot} is developing large computational/electronic load ride-through rules through NOGRR282/NPRR1308, motivated by observed DC and cryptomining load trips during normal voltage disturbances. When data centers are clustered in a region, the grid-side impact is aggravated. In the 2024 Virginia incident~\cite{virginia}, a single localized transmission fault propagated to data-center facilities across the region as a voltage sag, triggering their independent protection at once and amplifying the fault into a total load loss of roughly 1,500\,MW.

\begin{figure}[t]
    \centering
    \includegraphics[width=1\linewidth]{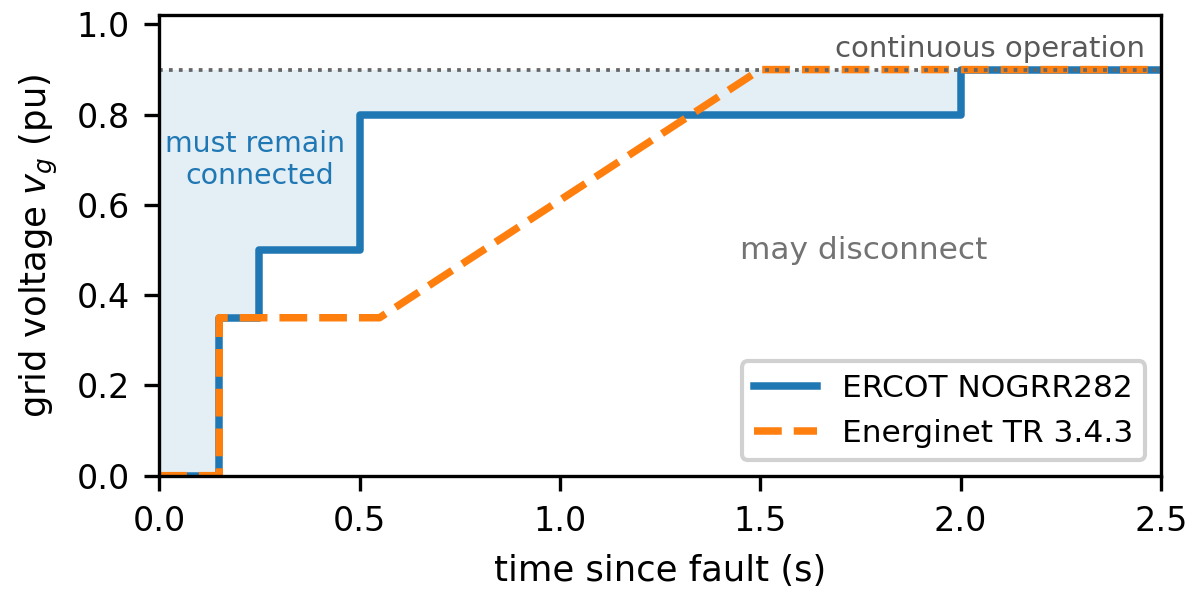}
    \vspace{-2em}
    \caption{Low-voltage ride-through envelopes of the two codes: ERCOT NOGRR282's band envelope (May 2026) and Energinet TR 3.4.3's voltage--time curve. }
    \vspace{-1.3em}
    \label{fig:gridcurves}
\end{figure}

These emerging codes share a three-part structure: the DC must stay connected while the voltage disturbance lies within a ride-through envelope (Figure~\ref{fig:gridcurves}), keep its active power draw above a floor during the sag, and restore its draw within a deadline after the sag clears. Codes differ only in the values that each part takes. For example, Energinet's VRT code~\cite{energinet}---the Danish code for transmission-connected demand facilities---places requirements on the same structure of three obligations as the ERCOT code, and the two codes only differ in the requirement values. Regarding how long the facility must stay connected during voltage sags or swells, ERCOT specifies voltage ranges with corresponding ride-through durations; in contrast, Energinet requires a piecewise-constant voltage--time curve. Regarding active power that the facility must keep drawing during the fault, ERCOT requires a level proportional to the sag for certain voltage ranges, while Energinet permits a floor of zero for all depths. Regarding recovery of active power draw after the fault, ERCOT sets a single target of $90\%$ of the pre-fault draw within two seconds, whereas Energinet stages three targets over $30$ seconds ($80/90/95\%$ at $5/20/30$\,s) and caps the overshoot.

VRT grid codes may require data centers to ride through both voltage sags and swells, but sags are the more challenging case. Upon sags, low-voltage ride-through (LVRT) involves all three requirements (connection, minimum power draw, recovery); but high-voltage ride-through (HVRT), in the example of ERCOT, imposes a connection requirement and an unchanged active power draw, making it an easier subset of the LVRT problem. 

Hence, we focus on LVRT in this paper,  although \sysname{}'s grammar can encode voltage swell and related power draw requirements as well and the synthesis query stays the same.
Unlike conventional LVRT requirements that are often specified for an individual generator or power-electronic device, data-center LVRT must be understood as a \textit{system-level property}. A data center consists of multiple interacting components, including UPS systems, battery energy storage, switchgear, power-electronics interfaces, internal distribution networks, and workload-dependent power consumption. 
In practice, during a voltage sag, tripping is initiated not by the computing equipment itself but by facility protection tuned to protect it. In the 2024 Virginia event, UPS control schemes that count voltage disturbances tripped the facility after repeated sags~\cite{virginia}.
LVRT analysis and certification for data centers is therefore fundamentally different from conventional device-level verification: it requires certifying that the aggregated data-center system satisfies grid-side ride-through requirements while respecting internal device constraints.

\subsection{Data Center Electrical and Compute Models}

From an electrical perspective, prior work models DC power delivery using UPS systems, BESS, switchgear, power-electronics converters, and internal distribution topology~\cite{xie2026enhancing}. These components make the load behavior different from a passive constant-power load through a combination of protection settings, converter limits, and reactive-power capability, which affects the data center's capability to comply with LVRT grid codes during a voltage disturbance. 
We adopt LinDistFlow~\cite{zhu2016fastlocal, baran1989optimal}, a standard linearized power-flow approximation for radial distribution networks, to model the internal network of the DC.
Specifically, the DC network extends from the grid interconnection point through transformers and distribution lines to internal buses, where compute loads and UPS and BESS converters are connected.
UPS and BESS resources are especially relevant for LVRT because they sit between the grid and the computing load~\cite{wang2014, aksanli2015}.
They can buffer active power, provide or consume reactive power, and protect internal equipment from voltage disturbances. Their behavior is constrained by hardware ratings: active-power limits, reactive-power limits, energy limits, and apparent-power capacity. In power-electronics models, active power $P$ and reactive power $Q$ often share the same converter capacity $S$, e.g., $P^2 + Q^2 \le S^2$.
These converter resources are typically what prior ride-through controller designs actuate: centralized and decentralized voltage control over the facility's internal distribution network~\cite{xie2026enhancing}, grid-forming control of UPS and BESS converters~\cite{shamseldein2025liability, abudyak2026mitigating}, passivity-based converter control for DC facilities~\cite{wang2026resilient}, and coordinated grid-side storage that absorbs load transients~\cite{kundu2025managing, mohammadi2026gridintegrationaidata}. 
These designs are validated on simulated faults, but treat the grid code informally: they can neither certify compliance nor establish that no compliant controller exists.

In terms of compute, data centers are increasingly dynamic loads. Prior work on GPU and AI workload power management shows that software can change the power profile through power caps, frequency scaling, scheduling, and workload shifting, for efficiency and demand response~\cite{choukse2025power, acun2026, liu2013, wierman2014}.
A related line coordinates data centers with the grid in normal operation, including voltage regulation via GPU-level actuation~\cite{liang2026gpu}.
These software knobs act on timescales too coarse for a low-voltage event, so they cannot serve as fault-time controls. The workload still impacts the ride-through problem through the pre-fault active power, which sets the grid code's power floor and recovery target.

\subsection{The Case for LVRT Formal Specifications}

Formal specifications are logical statements with precise semantics, in stark contrast to natural language, which is the common method for describing grid codes. 
Their precision is what enables automated verification and synthesis, and safety-critical domains have long adopted them for exactly this reason. 
Specifically, assume-guarantee specifications have been proposed as the algebra of compositional system design~\cite{benveniste2018contracts, incer2023algebraicaspectsassumeguaranteereasoning, Incer:EECS-2022-99}: one side states assumptions on its environment, and the other upholds guarantees whenever those assumptions hold. 
They have been applied to aircraft electric power systems~\cite{nuzzo2014contract}, space missions~\cite{pacti-space}, and networked control~\cite{chen2019compositional, 10549986}; this paper argues for taking a similar approach to LVRT analysis.

In particular, temporal logics support the expression of statements over signals that change over time.
Within the class of temporal logics, Linear Temporal Logic (LTL) and Metric Temporal Logic (MTL) operate over boolean propositions, while LVRT grid codes constrain real-valued signals against numeric thresholds under deadlines.
Signal Temporal Logic (STL) specifies requirements over exactly such real-valued signals in continuous time~\cite{maler2004monitoring}. Its atomic predicate is any Boolean-valued statement $\mu$ over the signals $x$ being considered. Its formulas combine predicates with Boolean operators and time-bounded temporal operators,
\begin{equation}
  \varphi \mathrel{::=} \mu(x) \mid \lnot\varphi \mid \varphi \land \psi
    \mid \varphi \lor \psi \mid \Box_{[a,b]}\,\psi
    \mid \Diamond_{[a,b]}\,\psi \mid \varphi\,\mathcal{U}_{[a,b]}\,\psi ,
  \label{eq:stl-grammar}
\end{equation}
where $\Box_{[a,b]}$ reads ``always on $[a,b]$,'' $\Diamond_{[a,b]}$ reads ``eventually within $[a,b]$,'' and $\varphi\,\mathcal{U}_{[a,b]}\,\psi$ reads ``$\varphi$ until $\psi$ within $[a,b]$''. Beyond the boolean verdict $x \models \varphi$, STL has a quantitative \emph{robustness} $\rho(\varphi,x)$: positive exactly when $x$ satisfies $\varphi$, with its magnitude being the margin of satisfaction or violation~\cite{fainekos2009robustness,bartocci}.
These operators can express many cyber-physical requirements, such as maintaining a signal within a safe range, responding within a deadline, or satisfying a condition over a time interval. LVRT has this same structure, since voltage ride-through, active power floor, and recovery are all time-dependent.
Synthesis from temporal-logic specifications, which decides whether any controller can satisfy a given specification, is studied extensively in robotics and control~\cite{belta2019formal, kress2018synthesis}. 
For STL in particular, synthesis can be formulated as an optimization problem~\cite{raman2014model,raman2015reactive}:  
Given a plant model $y = g(u, w)$ mapping a control signal $u$ and a disturbance $w$ to the output $y$, a feasible control set $K$, and an STL formula $\varphi$, synthesis solves
\begin{equation}
    u^{\star} = \arg\min_{u}\ J(u)
    \quad\text{s.t.}\;\rho\bigl(\varphi,\, y\bigr) \ge 0,
    \quad y = g(u, w),
    \quad u \in K,
    \label{eq:query}
\end{equation}
for a control cost $J$. The constraint $\rho \ge 0$ requires even the tightest clause of $\varphi$ at its tightest instant to hold. BluSTL~\cite{donze2015blustl} implements this method by encoding the bounded STL formulas as mixed-integer constraints over a discretized time window, solving~\eqref{eq:query} inside a receding-horizon loop and applying the first action each step, yielding a model predictive controller (MPC).
For power systems, this synthesis approach has targeted generation-side frequency regulation with energy storage~\cite{xu2017optimal, xu2019energy} and grid-supportive DER modes~\cite{zhang2019synthesizing}, but not load-side ride-through.
We build our synthesis on a Python reimplementation of BluSTL and instantiate~\eqref{eq:query} for our plant and specification in \S\ref{sec:algorithm}.

\lstdefinestyle{dslchunk}{mathescape=true, breaklines=true, basicstyle=\scriptsize\ttfamily,
  keywords={signal,param,let,event,first,after,at,assume,guarantee,in,if,then,Always,Eventually}, keywordstyle=\bfseries}

\section{\sysname Design}
\label{sec:system}
\begin{figure*}[t]
    \centering
    \includegraphics[width=0.85\linewidth]{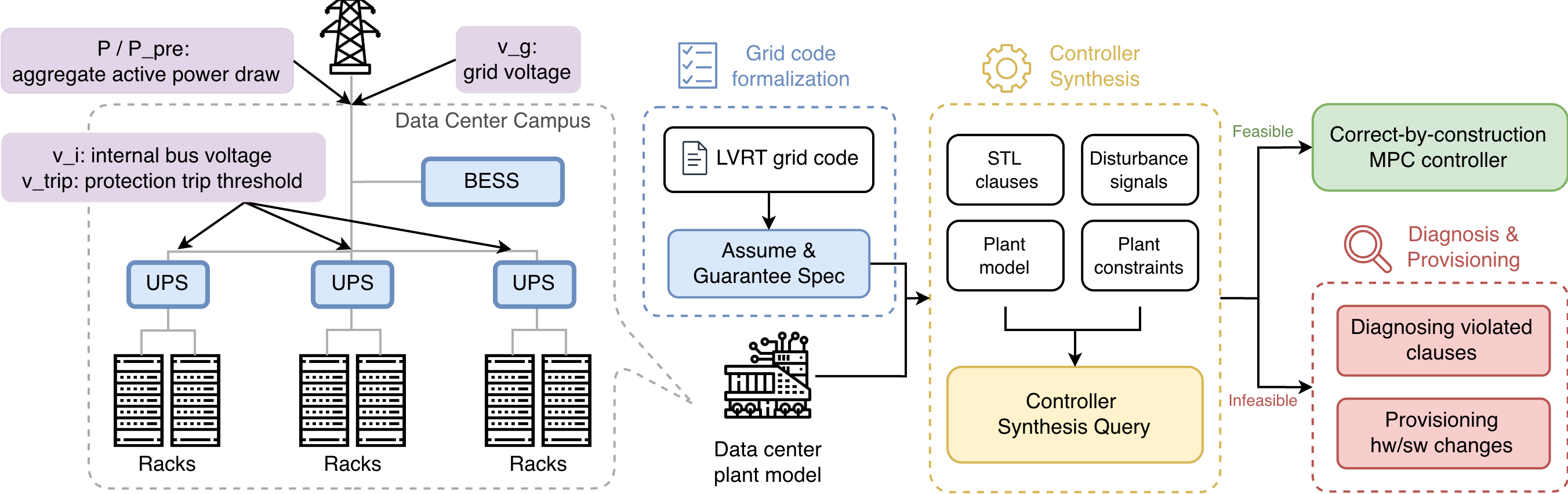}
    \vspace{-1mm}
    \caption{High-level architecture of \sysname, comprising three main components: grid code formalization (\S\ref{sec:dsl}), controller synthesis (\S\ref{sec:algorithm}), and diagnosis and provisioning (\S\ref{sec:provisioning}). A grid code is encoded as a formal spec in our STL-based DSL. The spec and the data center's parameters are encoded into a controller synthesis query. When feasible, the query returns a correct-by-construction MPC controller; when infeasible, diagnosis and provisioning computes changes that restore compliance. A simplified internal power topology of a data center is shown on the left, with LVRT signals at their measurement points.}
    \vspace{-0mm}
    \label{fig:sys_fig}
\end{figure*}

LVRT grid codes require data centers to stay connected during low voltage events.
To do so, a data center must carefully control its energy storage resources and UPS systems.
This section investigates how the data center should perform this control.
We first formalize grid codes using STL, which expresses LVRT requirements in a machine-readable format. We then provide a synthesis algorithm to find a controller for a given data center, deriving a control law and certifying its compliance with the grid code. 
If no such control law can be found, we perform diagnosis to suggest changes to the data center's parameters to achieve compliance---e.g., by increasing power electronic device capacity, redesigning the data center topology, or shaping workload power consumption.

\subsection{LVRT Grid Codes}
\label{sec:gridcode}
\begin{table}[t]
  \centering
  \small
  \begin{tabular}{@{}ll@{}}
    \toprule
    symbol & meaning \\
    \midrule
    $v_g$              & grid voltage at the point of connection \\
    $v_i$              & min internal bus voltage over $n$ compute buses \\
    $p$                & aggregate active power drawn from the grid \\
    $p_{\mathrm{pre}}$ & pre-fault active power (i.e., $p$ at the time of the fault) \\
    \bottomrule
  \end{tabular}
  \caption{Signals used to encode LVRT grid code.}
  \vspace{-0.5cm}
  \label{tab:signals}
\end{table}

Grid operators have introduced LVRT grid codes, which regulate the behavior of a data center during a voltage event and usually contain three components. The grid codes require the data center to stay connected during a grid voltage sag, keep active power draw above a level during the sag, and recover active power draw within a deadline after the sag is cleared. The following simplified example is provided by ERCOT~\cite{ercot}:

\vspace{3mm}
\begin{defn} [ERCOT LVRT grid code snippet]{defn:spec}
\vspace{-3mm}

A data center shall remain connected to the Transmission Grid during voltage conditions requiring ride-through. When grid voltage falls below 0.9 pu but remains above 0.8 pu, the data center shall continue consuming active power during the low voltage condition, and may reduce it proportional to the voltage drop, but shall return to 90\% of its pre-disturbance level within two seconds of voltage returning above 0.9 pu.
\end{defn}
\vspace{3mm}

LVRT involves the following three signals. \textit{(i)} $v_i$: In order for a data center to stay connected, its internal compute bus voltage $v_i$ must be above a threshold; otherwise, overly low voltages will trigger protection mechanisms that disconnect the data center from the grid and switch to a backup power source (e.g., diesel generators), in order to prevent damage to power electronics inside the data center. \textit{(ii)} $p$: The data center draws an aggregate active power $p$ from the grid; at the time of the fault, the power draw $p_{\mathrm{pre}}$ is called the pre-fault active power, read off the workload trace at the fault instant. This is the reference value, and it sets the scale of the power floor during the fault and the recovery target during recovery. \textit{(iii)}  $v_g$: The grid voltage $v_g$ at the point of data center / grid connection is the disturbance: this signal sags during a grid fault, and triggers an LVRT response from the data center.
Figure~\ref{fig:sys_fig} shows where these three signals are measured in the data center.
The duration for which the data center is required to stay connected to the grid depends on the severity of the sag. For more severe sags, the required duration is shorter; if the sag exceeds the threshold, the data center is allowed to instantaneously disconnect, and the grid code requirements are lifted. Table~\ref{tab:signals} lists the signals.

As \S\ref{sec:background} describes, LVRT codes share the same three obligations and differ only in the values each takes.
The next subsection introduces a language that the grid code author can use to turn these requirements into formal specifications.

\subsection{Formalizing LVRT Grid Codes} 
\label{sec:dsl}

As discussed in \S\ref{sec:gridcode}, the LVRT requirement is defined over multiple post-fault time segments, with time-coupled conditions and nontrivial rules for active-power floors during voltage sags and recovery. This temporal coupling and piecewise structure make it challenging to directly verify data center compliance using static operating-point analysis or conventional device-level checks. STL, on the other hand, has enough expressivity to write these requirements. Specifically, through the \emph{always} ($\Box$) and \emph{eventually} ($\Diamond$) operators over time intervals, it provides a convenient basis for the formalization of the LVRT requirements. Encoding LVRT requirements in STL not only removes ambiguity, but also enables automated synthesis of a controller for a given data center and grid code. Such a synthesized controller is \textit{guaranteed} to satisfy the requirements under all possible scenarios when such a controller exists. Next, we use the ERCOT example to illustrate this encoding in a new specification language for LVRT grid codes. Figure~\ref{fig:grammar} defines the language, and the rest of this subsection instantiates it for the ERCOT code.

A specification begins with variable declarations. Signals are time-varying quantities measured at the points shown in Figure~\ref{fig:sys_fig} (e.g., grid voltage $v_g$);  parameters are constant values for a data center (e.g., the equipment trip level $v_{\mathrm{trip}}$); variables are computed from signals and parameters. ERCOT declarations are as follows.

\begin{mdframed}[style=background]
\begin{lstlisting}[style=dslchunk, firstnumber=1]
signal Vg, Vi, p;  // definitions in Table 1
param  Vtrip;      // equipment tripping voltage
param  T_hold;     // recovery hold window
var    p_pre;      // power draw at time of fault
\end{lstlisting}
\end{mdframed}
\noindent Among these, $p_{\mathrm{pre}}$ and $T_{\mathrm{hold}}$ are interpretations we make where the ERCOT text is underspecified, rather than values taken literally from it; \S\ref{sec:discussion} details both.

An \textit{event} is defined when a signal of interest crosses a threshold. For instance, in the code below, a \texttt{fault} event occurs when $v_g$ falls below the required $0.90$\,pu, and a \texttt{recovery} event occurs when $v_g$ is brought back after a \texttt{fault}. The pre-fault power draw is taken from $p$ right before the fault event.  

\begin{mdframed}[style=background]
\begin{lstlisting}[style=dslchunk, firstnumber=last]
event fault     = fall(Vg, 0.90); 
event recovery  = rise(Vg, 0.90) after fault ;
p_pre  = p at fault; // sample just before fault
\end{lstlisting}
\end{mdframed}

\S\ref{sec:gridcode} describes the grid voltage sag ranges and the maximum durations for which the data center must stay connected. These are the conditions under which a data center is required to perform LVRT. The \texttt{assume} block captures them as follows:

\begin{mdframed}[style=background]
\begin{lstlisting}[style=dslchunk, firstnumber=last]
assume {
  var duration = {     // ride-through envelope
    Vg in [0.80, 0.90) => 2.0s;
    Vg in [0.50, 0.80) => 0.5s;
    Vg in [0.35, 0.50) => 0.25s;
    Vg in [0.00, 0.35) => 0.15s;
  }
  recovery - fault <= duration
}

\end{lstlisting}
\end{mdframed}

\noindent In the \texttt{assume} block, the ride-through envelope maps each sag depth to its maximum sag duration, and the final line assumes the grid voltage recovers within that duration.

If these assumptions hold, the data center is to uphold three LVRT guarantees shown in the code below. The \texttt{guarantee} block first defines the active power floor level $\mu$ that the grid code requires, then states the three clauses expressed in STL operators and anchored at the two events, \texttt{fault} and \texttt{recovery}.

\begin{mdframed}[style=background]
\begin{lstlisting}[style=dslchunk, firstnumber=last]
guarantee {
  // active power floor
  var mu = {
    Vg in [0.50, 0.90) => Vg * p_pre;
    Vg in [0.00, 0.50) => 0;
  }
  // clause 1: data center stays connected: 
  Always[fault, recovery] (Vi >= Vtrip);
  // clause 2: holds power floor: 
  Always[fault, recovery] (p >= mu);
  // clause 3: power draw eventually recovers 
  Eventually[recovery, recovery+2s](Always[0, T_hold] (p >= 0.9 * p_pre));
}
\end{lstlisting}
\end{mdframed}

\noindent The full requirement is the conjunction of three STL clauses, each a formula $\varphi$ of the grammar (Figure~\ref{fig:grammar}) with its time windows anchored at the declared events. We refer to them as connection ($\varphi_{\mathrm{conn}}$), active power floor ($\varphi_{\mathrm{flr}}$), and recovery ($\varphi_{\mathrm{rec}}$):
\begin{equation}
  \varphi := \varphi_{\mathrm{conn}} \land \varphi_{\mathrm{flr}} \land \varphi_{\mathrm{rec}}. 
  \label{eq:spec}
\end{equation}

\noindent \textit{Stay connected.}
The data center's internal voltage $v_i$ must stay above the power-electronics trip threshold $v_{\mathrm{trip}}$ over the fault window $[t_f,t_r]$:
\begin{equation}
  \varphi_{\mathrm{conn}} := \Box_{[t_f,\,t_r]}\,( v_i \ge v_{\mathrm{trip}} ). \label{eq:conn}
\end{equation}
\emph{Hold the power floor.} Throughout the fault duration, the active power must not fall below the floor level $\mu$ defined in the \texttt{guarantee} block:
\begin{equation}
  \varphi_{\mathrm{flr}} := \Box_{[t_f,\,t_r]}\,( p \ge \mu ). \label{eq:flr}
\end{equation}
\emph{Recovery.} After the recovery event, the active power must reach a target lower bound, $l = 0.9 \cdot p_{\mathrm{pre}}$, within a deadline $T=2$ seconds, and then hold that level for a window $T_{\mathrm{hold}}$. The \emph{eventually} operator $\Diamond$ over $[t_r,t_r+T]$, with a nested \emph{always} operator $\Box$ over $[0, T_{\mathrm{hold}}]$, captures this:
\begin{equation}
  \varphi_{\mathrm{rec}} := \Diamond_{[t_r,\,t_r+T]}\,\Box_{[0,\,T_{\mathrm{hold}}]}\,( p \ge l). \label{eq:rec}
\end{equation}
\noindent The code's text requires only the return of active power within the deadline after grid voltage recovers. We additionally require the level to hold for $T_{\mathrm{hold}}$, so that a momentary spike in power draw does not count as recovery; \S\ref{sec:discussion} discusses this interpretation in detail.

On the other hand, if the grid voltage goes beyond the envelope, the assumptions are violated and the data center does not need to perform LVRT.

\providecommand{\kw}[1]{\textbf{\texttt{#1}}}
\begin{figure}[t!]
  \centering
  \small
  \setlength{\tabcolsep}{3pt}
  \begin{tabular}{lcl}
    $spec$ & ::= & $decl^{+}$ \;\; $assume$ \;\; $guarantee$\\[3pt]
    $decl$ & ::= & \kw{signal} $x$ $\;\mid\;$ \kw{param} $x$\\
           & $\vert$ & \kw{var} $x$ $[\;=\, expr]$\\
           & $\vert$ & \kw{event} $e$ \,=\, (\kw{fall} $\mid$ \kw{rise})  ($x$, $const$) $[\kw{after}\ e]$\\[3pt]
    $assume$ & ::= & \kw{assume} \{ ($decl$ $\mid$ $pred$)$^{+}$ \}\\[3pt]
    $guarantee$ & ::= & \kw{guarantee} \{ ($decl$ $\mid$ $\varphi$)$^{+}$ \}\\
    $\varphi$ & ::= & $pred$ $\;\mid\;$ \kw{Always}$[I]\ \varphi$ $\;\mid\;$ \kw{Eventually}$[I]\ \varphi$\\
    $I$ & ::= & $[\tau,\ \tau]$ \qquad $\tau$ ::= $e$ $\;\mid\;$ $\tau + dur$ $\;\mid\;$ $dur$\\
    $pred$ & ::= & $expr$ ($\ge \mid \le \mid <$) $expr$ $\;\mid\;$ $expr$ \kw{in} $[\,const,\ const\,)$\\
    $expr$ & ::= & $const$ $\;\mid\;$ $x$ $\;\mid\;$ $\tau$ $\;\mid\;$ $x$ \kw{at} $e$ $\;\mid\;$ $expr$ ($+ \mid - \mid \cdot$) $expr$\\
           & $\vert$ & \{ ($pred$ $\Rightarrow$ $expr$)$^{+}$ \}\\
  \end{tabular}
  \vspace{-2mm}
  \caption{Grammar of the grid code specification language. Keywords are in \kw{bold}; $x$ stands for declared names, $e$ for declared events. \kw{signal}s are measured and \kw{param}s data-center-supplied, while \kw{var}s are derived; an \kw{event} binds the time instant of a threshold crossing. The \kw{assume} block describes the admissible disturbance; the \kw{guarantee} block is a conjunction of clauses $\varphi$: STL predicates over time windows $I$, which are event-anchored or relative. A predicate is a comparison or a band-membership test.}
  \vspace{-3mm}
  \label{fig:grammar}
\end{figure}

\subsection{Synthesizing LVRT Controllers}
\label{sec:algorithm}

\providecommand{\vgp}{v_g'}
Given an LVRT specification, our goal is to determine whether a data center can satisfy it. We address this task using controller synthesis. The data center exposes two signals, the internal bus voltage $v_i$ and the aggregate active power $p$. Neither can be controlled directly, as they are derived from the active and reactive power ($P$ and $Q$) that the BESS/UPS devices inject. We express the two converter control actions together as $u:= (P, Q)$, which is the control input for the data center.

To characterize how the control input $u$ affects the LVRT capability of data centers, we require a plant model of the data center's internal electrical network. We adopt the LinDistFlow model (\S\ref{sec:background}), under which the internal bus voltage $v_i$ is an affine function of the grid-side voltage and the converter injections $u := (P,Q)$. The active power $P$ contributes directly to the aggregate active power $p$ in the LVRT specification, while both $P$ and reactive power $Q$ affect internal voltage. These two actions are coupled by the converter apparent-power rating $S$, which imposes the constraint $P^2 + Q^2 \le S^2$. Thus, the reactive power used to support voltage reduces the remaining active-power headroom available to satisfy the active-power floor, and a converter with insufficient capacity may be unable to satisfy both requirements simultaneously. We denote these rating limits as the feasible control set $K$.

The controller synthesis procedure searches for control actions whose resulting $v_i$ and $p$ satisfy the specification discussed in \S\ref{sec:dsl}, or certifies that none exists.
A grid code like ERCOT's expresses the grid-voltage assumption as a set of bands, with each voltage band paired with a maximum ride-through duration. The \texttt{assume} block admits infinitely many disturbance traces, one for every sag shape inside the envelope, and the controller must satisfy the guarantees under all of them.  
Therefore, we synthesize against the single worst-case trace $\vgp$ that lower-bounds them all, holding each band at its lowest voltage for its full duration and ordering the bands from deepest to shallowest,

\begin{equation}
  \vgp(t) \;=\; \min\,\{\, v : \mathrm{dur}(v) \ge t \,\},
  \label{eq:wc-trace}
  \vspace{-0.1cm}
\end{equation}
where $\mathrm{dur}(v)$ is the maximum ride-through duration the specification's \texttt{assume} block assigns to a sag of depth $v$.
The result is the longest and deepest sag the assumption admits, the binding case for the connection clause; it is the lower boundary of the ERCOT region in Figure~\ref{fig:gridcurves}. When a code instead specifies a voltage--time curve, as Energinet does (the dashed curve in Figure~\ref{fig:gridcurves}), we use that curve directly as $\vgp$~\cite{energinet}.

The synthesis problem takes two disturbance signals, inputs the controller observes but does not choose: the grid voltage, for which we use the worst-case trace $\vgp$, and the workload's active power draw $w_{\mathrm{comp}}$, the compute demand the facility must serve. We write the two disturbances together as $w = (\vgp,\, w_{\mathrm{comp}})$. With the disturbances and the plant model in place, we instantiate the clauses in the specification. The voltage trace sets the fault and recovery times $t_f$ and $t_r$, turning each time window into a numeric interval. The workload trace sets the pre-fault draw $p_{\mathrm{pre}} = p(t_f)$, turning the active power floor $\mu$ and recovery target into numeric bounds. The plant model replaces the output signals $v_i$ and $p$ with their expressions in the control $u$ and the disturbance $w$. Substituting all three into the clauses of~\eqref{eq:spec} completes the instantiation, producing the concrete STL formula $\varphi$ over the control variables. Algorithm~\ref{fig:lowering-algo} in Appendix~\ref{app:lowering} lists the full procedure.

With the clauses $\varphi$, the plant model mapping $g$, the feasible set $K$, and the disturbance $w$ all concrete, we instantiate the synthesis problem~\eqref{eq:query} of \S\ref{sec:background}, with output $y := (v_i, p)$ and cost $J(u) = \lVert u \rVert$, selecting the least-effort compliant control $u^{\star}$ composed of per-step $P$ and $Q$. We solve the instantiated problem with our Python reimplementation of BluSTL (\S\ref{sec:background}). A feasible query yields an MPC controller that satisfies the instantiated guarantees for the specified voltage trace, workload trace, plant model, and hardware limits. An infeasible query means that no such controller exists for that synthesis instance, which we consider next.

\subsection{Diagnosis and Provisioning}
\label{sec:provisioning}

The control synthesis query in \S\ref{sec:algorithm} yields a binary result about the feasibility of synthesizing a control law that satisfies the grid code. In cases where such a controller does not exist, we further develop methods to diagnose the clauses of the grid code that cannot be met and compute the changes that could restore feasibility. We obtain both from the same synthesis query by treating the voltage sag depth and active-power floor as parameters of the specification and computing, for each sag depth, the largest floor at which the query stays feasible. 

\begin{figure}[t]
    \centering
    \includegraphics[width=1\linewidth]{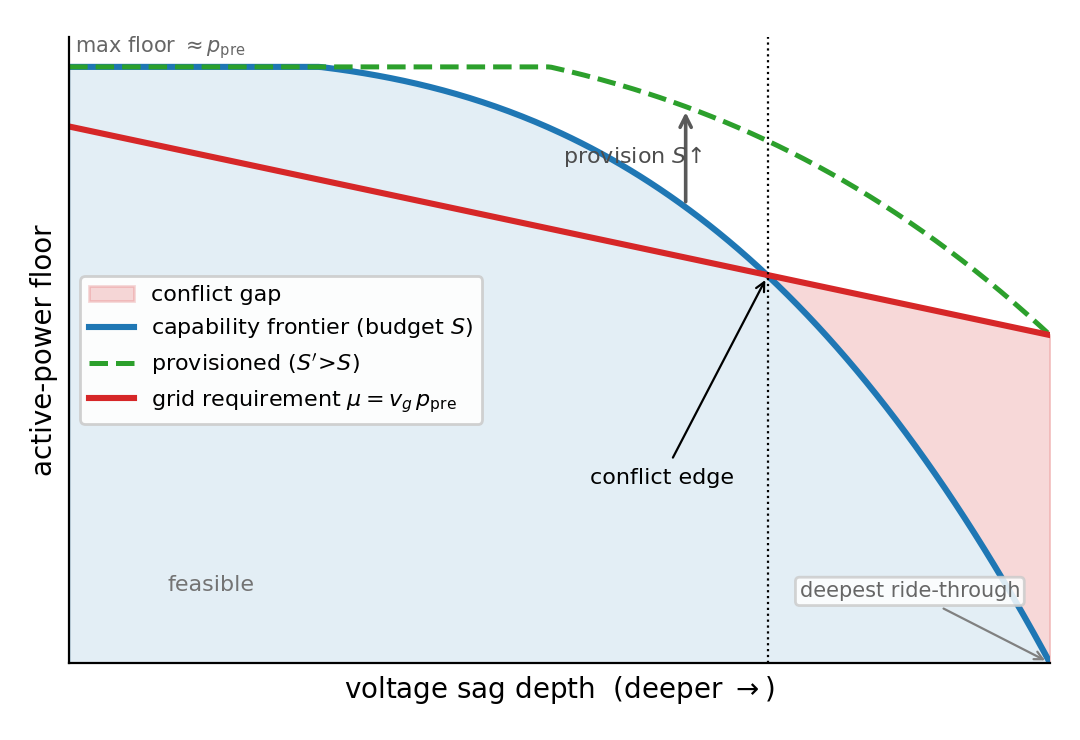}
    \vspace{-7mm}
    \caption{The \emph{conflict frontier.} The frontier (blue) is the largest active power floor the data center can sustain while staying connected at each sag depth. The grid requirement $\mu = v_g\,p_{\mathrm{pre}}$ (red) is the floor the code demands. Left of their crossing, the \emph{conflict edge}, the requirement is feasible; right of it the requirement exceeds the capability (solid red) and no controller exists. Provisioning (dashed) raises the frontier until the requirement lies inside the feasible region.}
    \vspace{-2mm}
    \label{fig:conflict-frontier}
\end{figure}

We call the resulting boundary the \emph{conflict frontier}, a new abstraction that we use for diagnosis, as shown in Figure~\ref{fig:conflict-frontier}. This is the maximum active power the data center can guarantee to draw from the grid while staying connected as a function of sag depth. This abstraction allows us to diagnose which grid code clauses cannot be jointly met.
Afterward, we modify the data center's parameters until the requirement lies fully inside the feasible region under the frontier.

\paragraph{The conflict frontier.} The frontier slopes downward because the connection clause and the active power floor clause draw on the same resource. To maintain the connection clause during a deeper voltage sag, the UPS's grid-facing converter needs to reserve a larger portion of its apparent-power rating for reactive-power support $Q$ to maintain the data center internal voltage $v_i$. Under the converter rating constraint $P^2 + Q^2 \le S^2$,  this increased reactive-power provision reduces the remaining capacity available for active-power $P$. Consequently, the largest floor the converter can sustain decreases as the sag depth increases. The requirement $\mu = v_g \cdot p_{\mathrm{pre}}$ also decreases with depth, so the most challenging sag is not necessarily the deepest one; the binding depth is wherever the requirement exceeds the frontier the most.

\paragraph{Computing the frontier.} The active power floor for each voltage sag depth is monotone: a lower floor is always easier to sustain since it requires lower $P$ and frees the converter headroom for $Q$. Leveraging monotonicity, we locate the largest sustainable active power floor for each voltage sag depth using binary search. Sweeping the voltage sag depth traces the frontier, the largest sustainable floor at each depth. The grid code's requirement is the second curve over the same axes, where the code demands the active power floor $\mu = v_g \cdot p_{\mathrm{pre}}$ at each depth. When the requirement exceeds the frontier, no controller exists, and the crossing marks the \emph{conflict edge}, the deepest sag the data center can both ride through and hold the required floor. We compute the frontier with inspiration from two established directions of work. Parametric STL treats values inside a temporal-logic spec as parameters~\cite{asarin2011parametric}, often used in specification mining to derive the tightest spec a fixed model satisfies~\cite{jin2015mining, hoxha2017mining}; we likewise treat the active-power floor and the voltage sag depth as parameters of the STL spec, but the inner oracle at each parameter value is whether a controller~\eqref{eq:query} exists, instead of whether a fixed model satisfies the formula. Multiparametric programming characterizes how the solution of an optimization varies with its parameters~\cite{bemporad2002explicit}, used in explicit model predictive control to precompute the control law as a function of state; we likewise sweep these parameters, but keep only the boundary of the feasible region as the conflict frontier, instead of the control law across it.

\paragraph{Diagnosis.} To attribute an infeasibility to specific clauses, we recompute~\eqref{eq:query} over subsets of the guarantee clauses and isolate the minimal infeasible subset, which is the set of clauses for which no controller exists but dropping any single clause makes the rest feasible. This is the Irreducible Infeasible Subsystem (IIS) of the program~\cite{chinneck1991locating}. IIS extraction usually explains an infeasible program to the author; in our case, we leverage IIS to identify the set of grid code clauses in conflict, turning the infeasibility reported by the solver into a diagnosis.
With three clauses, as in the ERCOT example, every subset can be tested with ease. If a grid code presents more guarantee clauses, we can rely on the MILP solver to extract an irreducible infeasible subset on its own~\cite{gurobi}. A clause that is infeasible on its own cannot be satisfied with any controller. Clauses that are feasible individually but infeasible together are in conflict, such as the connection and active power floor clauses competing for the apparent power rating $S$.

\paragraph{Provisioning.} Provisioning closes the gap between the requirement and the frontier by raising the frontier or lowering the requirement through three knobs. 1) The UPS/BESS power-conversion-system (PCS) apparent-power rating $S$ raises the frontier directly: a larger apparent power rating allows more $P$ and $Q$ simultaneously, with more reactive power headroom to maintain internal voltage and more active power to maintain the active power floor. 2) As part of the plant model, increasing the interconnection reactance of lines and transformers raises the converter's voltage authority: a larger reactance gives more internal-voltage change per unit of reactive power injection, so lower $Q$ holds the internal voltage at a given sag depth and more of the rating $S$ is left for $P$. 3) Adjusting the workload power profile acts on the requirement: lowering the pre-fault draw $p_{\mathrm{pre}}$ lowers the active power floor requirement during the fault, lowering the grid code requirement curve against the conflict frontier. Each of the three knobs is monotone, allowing us to apply binary search to derive the value at which the query first becomes feasible, i.e., the smallest change that restores compliance.

\section{Results}
\label{sec:results}
The evaluation answers three key questions: (a) on a live grid, does the synthesized controller comply with the grid code where alternative control schemes fail (\S\ref{sec:eval-live}); (b) when the synthesis query is infeasible, can \sysname{} diagnose which clauses conflict or whether any individual clause is unsatisfiable (\S\ref{sec:eval-diagnosis}); and (c) what is the smallest hardware or workload change that restores compliance (\S\ref{sec:eval-provisioning})?

\paragraph{Setup.} We evaluate on a data center instantiated from the Vulcan facility data~\cite{vaidhynathan_vulcan_2025}: eight compute nodes, each modeled as a 20\,MW load supplied through an upstream UPS with a 25\,MVA grid-facing converter; two 20\,MVA battery (BESS) units sit at the facility's distribution bus; and a 40\,MW cooling plant brings the rated facility draw to 200\,MW. The compute draw follows the measured power of a synchronous AI training job~\cite{choukse2025power}, normalized to peak and scaled to the facility's compute capacity.

\paragraph{Live grid simulator.} The controller of \S\ref{sec:algorithm} is synthesized against the worst-case trace of the grid code envelope; \S\ref{sec:eval-live} runs it closed-loop on a grid simulator, which changes its input in two ways. First, the measured voltage is one realized transient inside the assumed envelope, produced by the grid model, not the worst-case bound of~\eqref{eq:wc-trace}. Second, the loop is closed: the data center's own $P$ and $Q$ enter the grid's power-flow solution and change the grid voltage the data center measures next. We connect the data center to ANDES~\cite{cui2021andes} on the 140-bus NPCC system on bus 128. The disturbance is a three-phase fault near the interconnection, applied at $t{=}1$\,s and held for its band's ride-through duration before clearing ($2$\,s for a sag in $[0.8, 0.9)$, $0.5$\,s for sags in $[0.5, 0.8)$). The closer the fault, the deeper the interconnection voltage sags; we vary the fault impedance on the transmission grid to simulate faults at different electrical distances, producing voltages of different depths inside the envelope the code assumes. Every $0.02$\,s the controller reads the interconnection voltage, re-solves with the measurement held constant over its horizon, converts the first action to aggregate $P$ and $Q$, and writes them to the load; ANDES takes the $P$ and $Q$ and advances, and the next interconnection voltage measurement closes the loop. In every run, the facility's undervoltage protection disconnects the data center if the internal voltage $v_i$ remains below $v_{\mathrm{trip}} = 0.9$ for $0.02$\,s, and a disconnected facility draws zero power.

\paragraph{Experiment design.} The subsections map to the three questions and isolate one variable at a time. To answer (a), \S\ref{sec:eval-live} varies the controller on a provisioned variant of the facility with converters rated at $S{=}40$\,MVA per node, large enough that a compliant controller exists at every voltage sag depth the grid code covers: no controller and a penalty-based controller from prior work~\cite{xie2026enhancing} serve as baselines against the controller synthesized from~\eqref{eq:query}, each run with the live grid simulator on the same faults. To answer (b) and (c), \S\ref{sec:eval-diagnosis} and \S\ref{sec:eval-provisioning} hold the synthesized controller fixed and return to the as-built $25$\,MVA rating, where the synthesis query is infeasible and the diagnosis and provisioning of \S\ref{sec:provisioning} take over.

\subsection{Ride-through Effectiveness on a Live Grid}
\label{sec:eval-live}

\begin{figure}[t]
    \centering
    \includegraphics[width=1\linewidth]{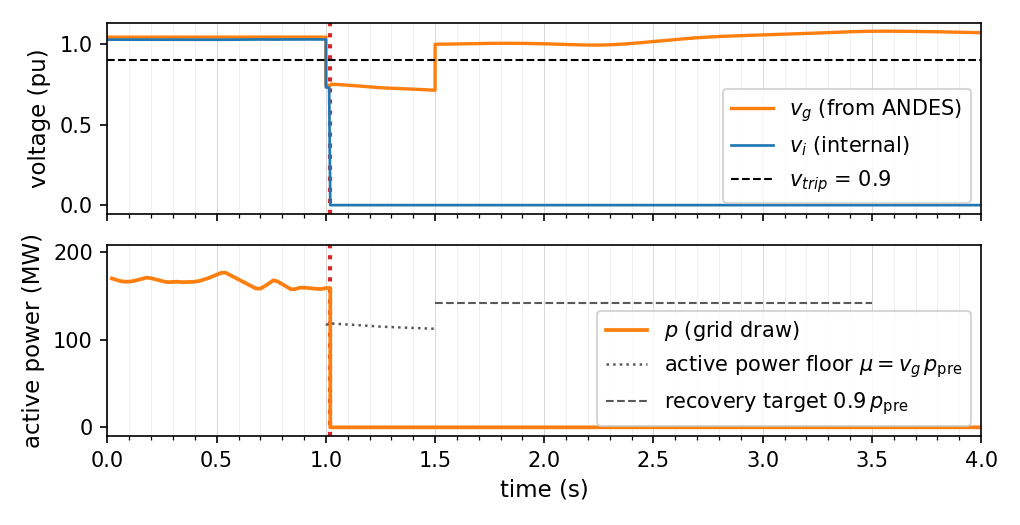}
    \vspace{-2.5em}
    \caption{No-controller baseline. The fault at $t{=}1$\,s trips the data center; both the internal voltage and the active power draw drop to zero, violating all three clauses.}
    \vspace{-0.1cm}
    \label{fig:f1}
\end{figure}

\paragraph{No controller (Figure~\ref{fig:f1}).} With the UPS and BESS taking no control action to inject active or reactive power, the internal voltage follows the grid voltage. The fault sags $v_g$ to $0.71$ and $v_i$ follows; after the $0.02$\,s pickup the protection disconnects the facility. All three clauses fail together: $v_i$ drops below $v_{\mathrm{trip}}$ and the data center disconnects (connection); as a result, the $160$\,MW draw drops to zero, below the ${\approx}110$\,MW floor (floor), and it never returns to $0.9\,p_{\mathrm{pre}}$ within the $2$\,s deadline (recovery). The grid recovers after the fault is cleared and $v_g$ settles above its pre-fault value: the disconnection becomes a disturbance. At one facility the post-fault voltage overshoot is modest, but it becomes a systemic event when a cluster of facilities trips at the same time, as observed in the Virginia incident in 2024~\cite{virginia}, the behavior the code is written to prevent.

\begin{figure}[t]
    \centering
    \includegraphics[width=1\linewidth]{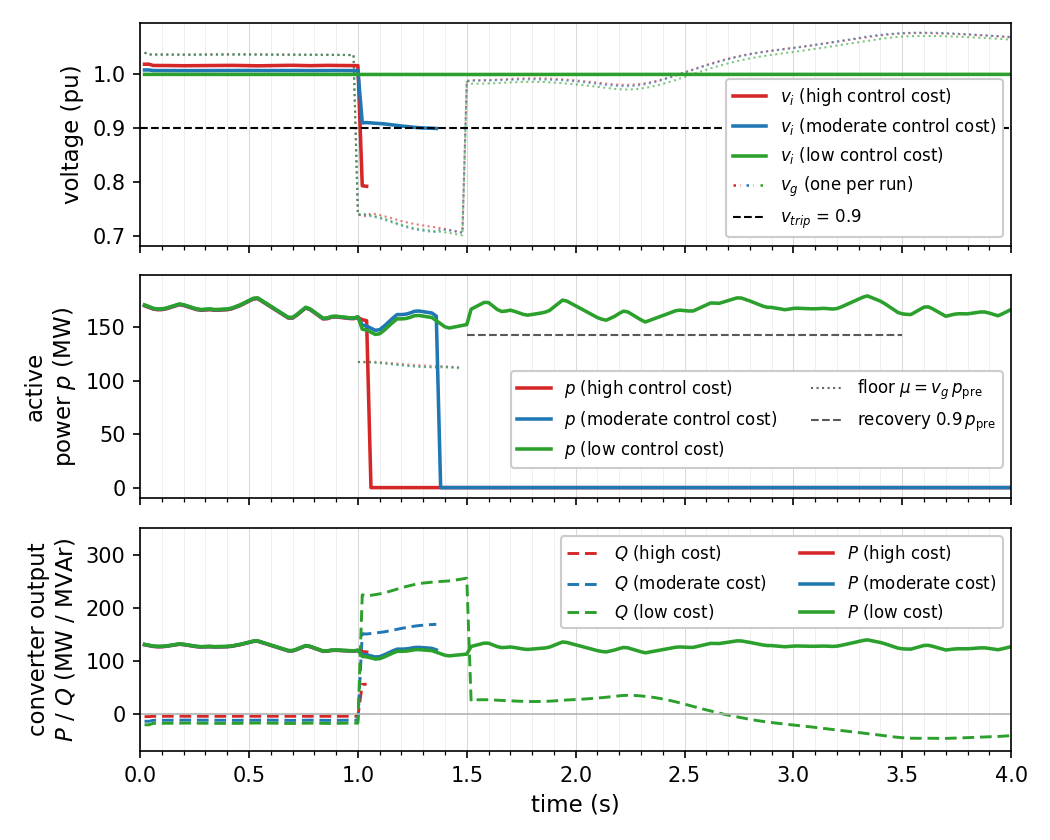}
    \vspace{-0.5cm}
    \caption{Penalty-based controller under three penalty weights on the same fault. The grid voltage $v_g$ differs across runs because the loop is closed. Controller with high cost trips data center at the fault onset; one with moderate cost holds $v_i$ near $0.90$ and trips mid-fault at $t{=}1.34$\,s; one with low cost rides through.}
    \vspace{-0.3cm}
    \label{fig:f2}
\end{figure}

\paragraph{Penalty controller (Figure~\ref{fig:f2}).} Penalty-based voltage control is the standard alternative, which can hold the internal voltage to avoid tripping, but compliance depends on a hand-tuned weight and carries no guarantee. We implement a data-center voltage controller from prior work~\cite{xie2026enhancing}: at each step, the control law chooses the converter's P and Q injections that pull the internal voltage toward nominal, with a weight $c$ penalizing deviation from normal operation. It's expressed as $\min_u\, \lVert v_i - \mathbf{1} \rVert^2 + c\,\lVert u - u_{\mathrm{nom}} \rVert^2$ over the same feasible set $K$, where $u_{\mathrm{nom}}$ is normal operation: the converters simply import compute draw and provide no reactive support. This is the objective of the controller that Xie et al. used~\cite{xie2026enhancing}, written in our notation and with their per-device effort weights collapsed to the single scalar $c$ and their box bounds replaced by $K$, so the penalty controller commands the same hardware as the synthesized one. In such a penalty-based controller, no grid code clause of $\varphi$ appears in the objective. We run three weights on the same fault as the no-controller run. At $c{=}5$ (high control cost) the controller is reluctant to act, with $v_i$ reaching only $0.79$ and the data center tripping at the fault onset. At $c{=}0.001$ (low control cost) it holds $v_i \approx 1$ and the active power draw stays above the floor as a result, which is grid code compliant. At $c{=}0.8$ (medium control cost), the controller holds $v_i \approx 0.90$ with almost no margin through the first $0.3$\,s of the fault, then $v_i$ slips under the threshold and the data center trips at $t{=}1.34$\,s; the active power draw drops to zero, and the floor and recovery clauses fail with it. Compliance thus depends entirely on the hand-picked weight, and the formulation gives no prior indication of the compliant range: a weight that survives the fault onset can fail partway through, and the floor and recovery are met only incidentally. In our case, the synthesis query~\eqref{eq:query} decides compliance offline before the fault instead of during it: it either returns a compliant controller by construction, or reports infeasibility at design time with no compliant controller.

\begin{figure}[t]
    \centering
    \includegraphics[width=1\linewidth]{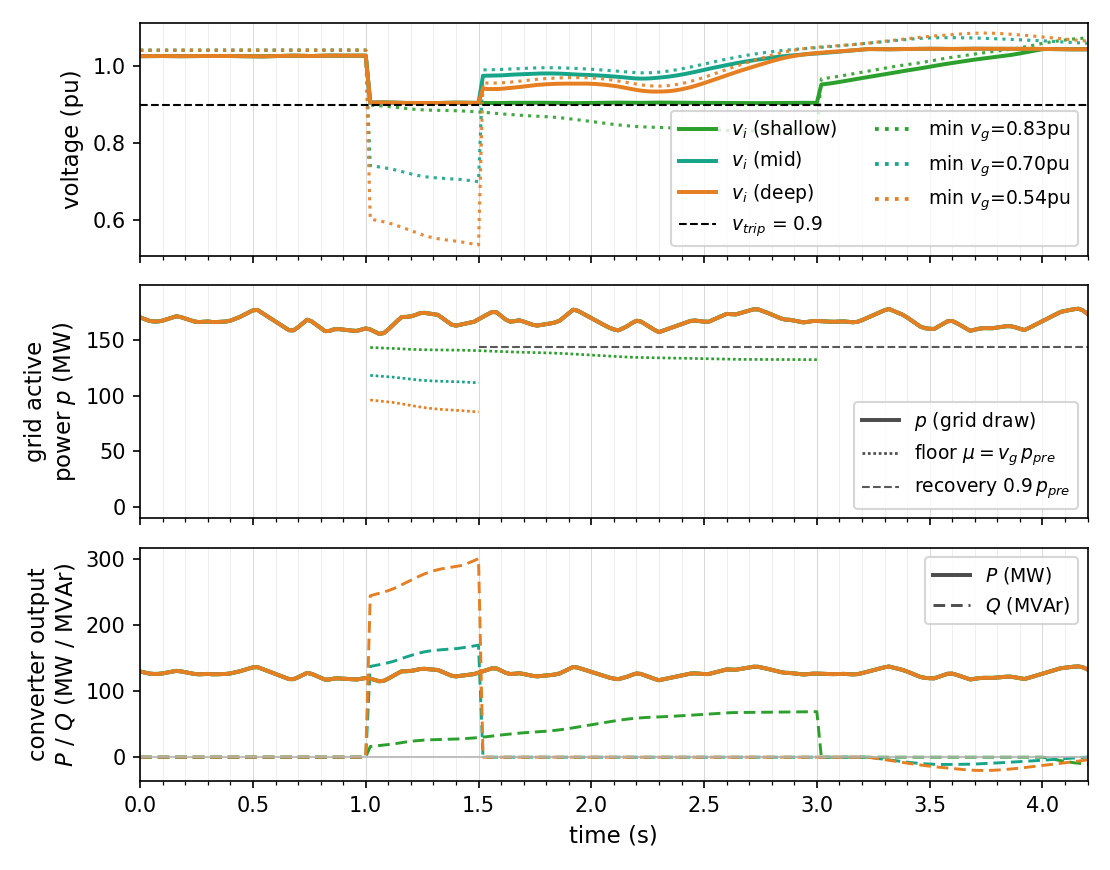}
    \vspace{-0.5cm}
    \caption{Synthesized controller on three sag depths. Thin dotted curves in the top panel are each run's grid voltage $v_g$; the shallow sag is held for its band's $2$\,s duration and the mid and low sags for $0.5$\,s, all inside the ERCOT envelope. In every run the controller holds $v_i$ just above $v_{\mathrm{trip}}$ while the draw $p$ tracks the workload above its floor $\mu$.}
    \vspace{-0.5cm}
    \label{fig:f3}
\end{figure}

\paragraph{Synthesized controller (Figure~\ref{fig:f3}).} The controller synthesized from~\eqref{eq:query} rides through three voltage sag depths inside the assumed voltage envelope and complies with the grid code. During each fault, the controller holds $v_i$ slightly above $v_{\mathrm{trip}}$ (a $0.005$\,pu robustness margin), injecting more reactive power the deeper the sag: roughly $70$, $170$, and $300$\,MVAr at depths $0.83$, $0.70$, and $0.54$. The converter active power simply tracks the compute demand and meets the floor with no curtailment, because at this rating the reactive support and the full compute load fit inside the apparent-power headroom together, so compliance with active power floor as well as recovery costs no active power control action. The guarantee is conditional on exactly this headroom, where the facility's $40$\,MVA converters sit above the minimum that \S\ref{sec:eval-provisioning} derives; on the as-built $25$\,MVA rating, the $P$ and $Q$ no longer fit together and the query returns infeasible and no controller exists. The next subsections turn that infeasibility into a diagnosis and a provisioning decision.

\subsection{Infeasibility Diagnosis}
\label{sec:eval-diagnosis}

This subsection and the next fix the controller and return to the baseline data center with the as-built $25$\,MVA rating in the Vulcan system~\cite{vaidhynathan_vulcan_2025}. The question is design-time feasibility, so no grid simulator is involved. The synthesis query of \S\ref{sec:algorithm} takes three inputs. The specification is the ERCOT code of \S\ref{sec:dsl}. The plant model is the Vulcan facility at its as-built rating, modeled with LinDistFlow. The grid-voltage disturbance is the worst-case envelope of~\eqref{eq:wc-trace}, restricted to the depths where the active power floor applies ($v_g \ge 0.5$); below $0.5$ the code releases the active power floor and permits momentary cessation, the facility briefly stopping its grid power draw and restoring it once the voltage returns, so no converter action is required during the fault. The workload disturbance is the measured trace with the grid fault aligned to the trace peak. A fault can land anywhere on the trace, and $p_{\mathrm{pre}}$, the active power draw at the fault instant, sets the floor $\mu = v_g\,p_{\mathrm{pre}}$, so this alignment is the worst case, $p_{\mathrm{pre}} = 200$\,MW. The synthesis query returns infeasible for this combination of grid code spec, plant model, and disturbance, meaning no controller satisfies $\varphi$.

\begin{figure}[t]
    \centering
    \includegraphics[width=1\linewidth]{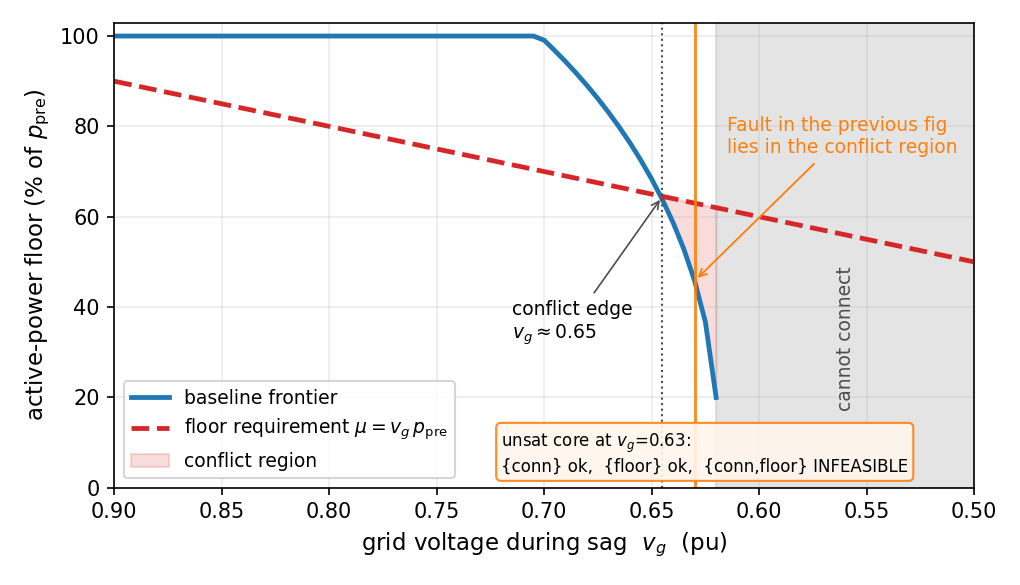}
    \vspace{-2.5em}
    \caption{Conflict frontier of the as-built facility, swept at $81$ depths ($0.005$\,pu apart). The frontier is capped at $100\%$ of $p_{\mathrm{pre}}$, since a floor above the pre-fault draw is meaningless. The orange marker at $v_g = 0.63$ is the depth examined in Figure~\ref{fig:conflict-trace}, with its clause-subset diagnosis in the box.}
    \vspace{-3mm}
    \label{fig:frontier-eval}
\end{figure}
\paragraph{The conflict frontier (Figure~\ref{fig:frontier-eval}).} The synthesis against the as-built facility fails in two ways, requiring different fixes. At moderate voltage sag depths, the connection and the active power floor clauses conflict: each clause is satisfiable alone, but their conjunction is not (the minimal infeasible subset of \S\ref{sec:provisioning}). At deeper voltage sags, connection is infeasible alone. Recovery is never among the failing clauses. We locate the two modes with the conflict frontier of \S\ref{sec:provisioning}, finding at each depth between $0.90$pu and $0.50$pu the largest floor the data center can sustain while staying connected, and re-solving the query over clause subsets at each depth. The required floor crosses the frontier at $v_g \approx 0.65$pu, the conflict edge. Between $0.65$pu and $0.62$pu the data center can either stay connected or maintain the active power floor, but not both. Below $0.62$pu, holding internal voltage to stay connected alone is infeasible, at what we call the connection wall. 
At the wall the frontier ends at $40$\,MW ($20\%$ of $p_{\mathrm{pre}}$) rather than zero, where the converters are fully spent on reactive support, and the residual is the cooling draw, which is served without a converter and sits outside the control knobs we model.
Inside the conflict region, raising the capability or lowering the requirement both restore feasibility; below the wall, only raising the capability can help.

\begin{figure}[t]
    \centering
    \includegraphics[width=1\linewidth]{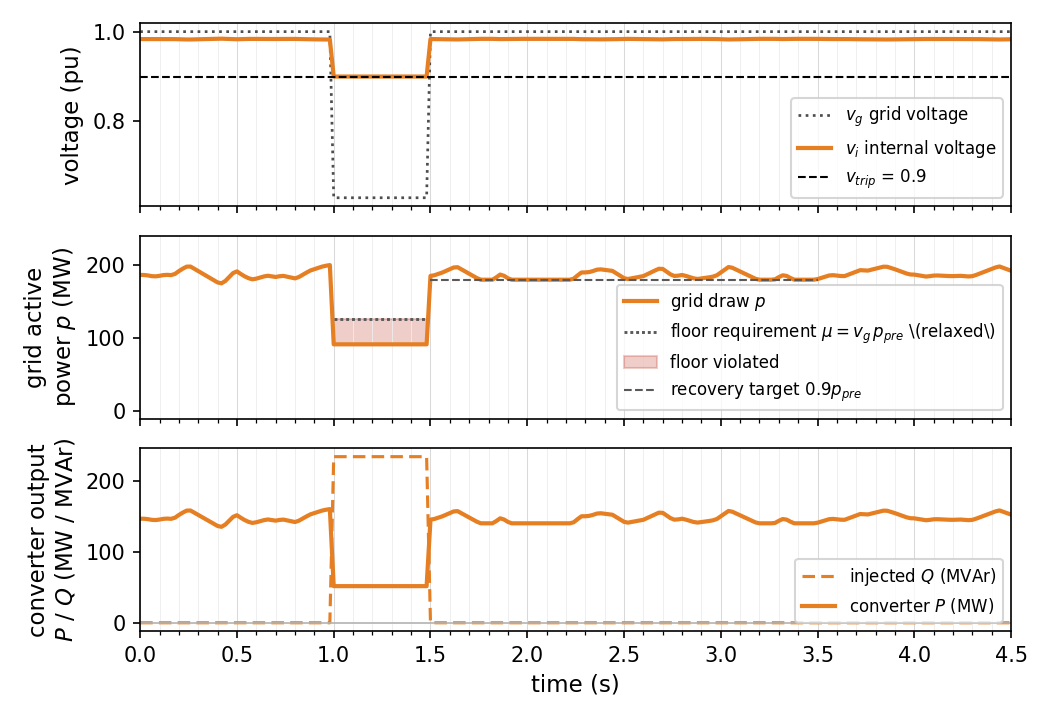}
    \vspace{-2em}
    \caption{The conflict at $v_g = 0.63$. Only the connection clause is enforced for demonstration; with best effort on the active power clause, the converter headroom is not enough for active power to satisfy the floor clause, causing a violation.}
    \vspace{-5mm}
    \label{fig:conflict-trace}
\end{figure}

\paragraph{The conflict on the time axis (Figure~\ref{fig:conflict-trace}).} Figure~\ref{fig:conflict-trace} shows what the conflict means in the time domain: even the largest active power draw the facility can sustain while staying connected falls short of the floor the spec requires. An infeasible query returns no controller, so there is no control action trajectory to see the violation. We therefore plot a relaxation instead, by keeping enforcement on the connection clause and making the best effort toward the active power floor clause. Staying connected takes $234$\,MVAr of reactive power $Q$, which nearly fills the converter headroom $S$, and the remaining active power $P$ supports an active power draw of only $92$\,MW, $34$\,MW below the $126$\,MW floor requirement. In other words, the connection clause is only satisfiable by giving up the active power floor clause inside the conflict region.

\subsection{Provisioning}
\label{sec:eval-provisioning}

For a provisioning team, this subsection answers the question of which hardware to specify so that the facility is capable of complying with the LVRT grid code. Each knob is monotone, so a binary search on the synthesis query returns the smallest value at which it turns feasible, a design requirement derived directly from the grid code and available before the facility is built or upgraded.

\begin{figure}[t]
    \centering
    \includegraphics[width=1\linewidth]{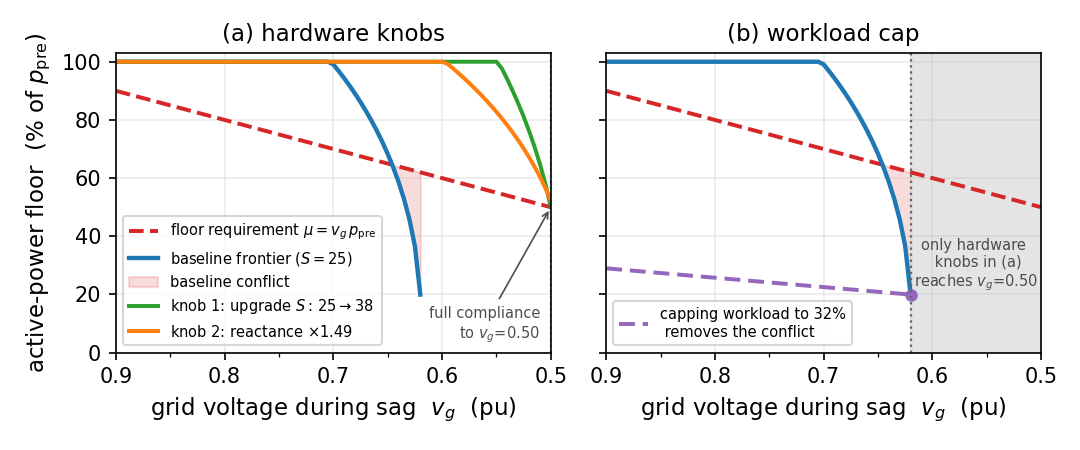}
    \vspace{-5mm}
    \caption{Provisioning. (a) Hardware: each knob is plotted at the minimal value that restores compliance, the converter rating $S$ raised from $25$ to $38$\,MVA or the interconnection reactance scaled by $1.49$; either lifts the frontier above the requirement at every depth down to $v_g = 0.50$. (b) Workload: at the minimal cap of $32\%$ of rated draw, the capped requirement (dashed) falls below the frontier at every depth where the facility can connect, removing the conflict, while compliance is still bounded by the connection wall.}
    \label{fig:provision-hw}
    \label{fig:provision-sw}
    \vspace{-3mm}
\end{figure}

\paragraph{Provisioning, hardware (Figure~\ref{fig:provision-hw}a).} Provisioning closes the gap between the requirement and the capability. \S\ref{sec:provisioning} exposes two hardware knobs and they help raise the frontier: the UPS converter rating $S$ and the interconnection reactance. A larger $S$ widens the eight per-node apparent power rating headroom with more $Q$ for connection and more $P$ for the floor at the same time. A larger reactance moves more internal voltage per unit of $Q$, so less of each headroom holds $v_i$ and more remains for $P$. Both are monotone and bisection returns the minimal change, $S$ from $25$ to $38$\,MVA per node ($+50\%$) or reactance scaled by $1.49$ in the model. Both hardware knobs lift the frontier above the requirement at every depth down to $v_g = 0.50$, the bottom of the active power floor band. Both knobs also push the connection wall below $0.50$. The BESS apparent power rating is a separate yet weaker knob, so we don't demonstrate in Figure~\ref{fig:provision-hw}. Because BESS reactive power reaches the internal buses only through the facility trunk, its reactive power injection barely helps maintain internal voltage to avoid tripping. 

\paragraph{Provisioning, workload (Figure~\ref{fig:provision-sw}b).} The workload enters this analysis only through the pre-fault draw. During a fault the UPS battery carries the compute, which costs little energy over events of a few seconds, and the grid-side draw becomes a control that the synthesis sets freely, up to the converter rating. What the workload does during the fault therefore never appears in the fault-time constraints, but only as $p_{\mathrm{pre}}$, the active power draw at the fault instant, which sets the requirement $\mu = v_g\,p_{\mathrm{pre}}$. The software knob caps the pre-fault workload. The cap lowers $p_{\mathrm{pre}}$ and the requirement with it and leaves the frontier where it is. In our model, capping the draw at $32\%$ of rated ($p_{\mathrm{pre}} = 64$\,MW) brings the requirement below the capability at every depth the data center can connect, with no hardware change, removing the $\{\varphi_{\mathrm{conn}}, \varphi_{\mathrm{flr}}\}$ conflict entirely. However, the connection wall does not move with software changes alone, as it stays constrained by the maximum reactive power the UPS hardware can inject; no software capping connects the data center below $0.62$ and compliance down to $0.50$ stays infeasible. 

The contrast summarizes the provisioning section. Hardware raises data center capability by moving both the frontier and the connection wall to reach full grid code compliance. A workload cap only lowers the grid code floor requirement and is still bounded by the hardware constraints. Even a facility capped to a third of its rated power draw clears only the conflict band. Hardware provisioning is the more fundamental fix, while workload capping alone is only a stop-gap for sags above the connection wall.

\section{Discussion}
\label{sec:discussion}

\paragraph{Grid code ambiguity.} Formalizing ERCOT's grid code exposed two underspecified terms. First, the code scales its active power floor and recovery requirements to the facility's ``pre-disturbance level'' but does not define how that value is read. Over a fluctuating AI workload, different methods of deriving the ``pre-disturbance level'' could give drastically different values.
We take the instantaneous active power draw (\S\ref{sec:gridcode}) for this work, instead of taking an average over some window. Second, the recovery clause requires the facility to ``return to 90\% of its pre-disturbance level within two seconds,'' but presumably, a facility that touches the level at the deadline and drops off immediately still violates the code. We add the hold window $T_{\mathrm{hold}}$ of~\eqref{eq:rec} so that recovery means genuine restored consumption rather than a momentary spike. These nuances emerged only when translating the natural-language grid code into a formal specification. This highlights the value of a formal approach, as it forces ambiguous requirements to surface and be made explicit before they are enforced.

\paragraph{Modeling state-of-charge.} We do not model energy storage state of charge. Batteries in UPS and BESS may charge or discharge during ride-through. On the discharging side, riding-through a voltage event spends a small fraction of the stored energy, since a VRT event lasts only a few seconds, while the UPS's storage capacity is provisioned to carry the full compute load for tens of seconds and the BESS for minutes. On the charging side, it requires batteries not held at full charge, and the headroom needed is likewise a small fraction of their provisioned capacity. Thus, we do not model the state-of-charge (SoC), and assume the UPS and BESS have both enough stored energy for discharging and enough headroom for charging during a VRT event.

\paragraph{Beyond one code and one facility.} Similar to LVRT grid codes, other grid codes regulate data center behavior for other types of grid disturbances (e.g., frequency ride-through). ERCOT writes frequency ride-through in the same document as voltage ride-through with the same structure: an assumption envelope on grid frequency, a power draw requirement dependent on the pre-disturbance level, and a power draw recovery clause. \sysname has the potential to solve frequency ride-through with minimal changes: our specification language can express these clauses with the same shape, just with frequency declared as the disturbance signal instead; however, the plant model and control knobs differ, and we leave the adaptation to future work. 
On the data center side, while \sysname certifies individual data centers' compliance with the LVRT grid code, the aggregated behavior of clustered data centers and their impact on the grid remains an open grid-side question.

\section{Conclusion}

Data centers have grown into loads large enough that their protective disconnection threatens the grid during voltage disturbances, and emerging VRT grid codes now require them to ride through such voltage events. This paper solves data center compliance with VRT grid codes using a controller synthesis approach. \sysname{} expresses a grid code as an assume-guarantee specification in Signal Temporal Logic, synthesizes a controller that complies by construction whenever one exists, and turns infeasibility into a conflict frontier, a diagnosis of the conflicting clauses, and a provisioning recommendation for the smallest hardware or workload change that restores compliance. On a model of a 200\,MW facility in a closed loop with a transmission grid simulator, the synthesized controller rides through faults that trip the uncontrolled facility, and the diagnosis traces the as-built rating's infeasibility to converter capacity, for which a hardware upgrade restores feasibility. As grid operators write ride-through rules for large loads and data center operators adapt to comply, \sysname{} gives both sides a machine-checkable interface between grid requirements and data center capabilities.

\bibliographystyle{ACM-Reference-Format}
\bibliography{reference}

\appendix

\section{The Encoding Procedure}
\label{app:lowering}

Algorithm~\ref{fig:lowering-algo} lists the full encoding procedure of \S\ref{sec:algorithm}. It constructs the worst-case trace from the specification's \texttt{assume} block, resolves the event times, instantiates the declared variables and the guarantee clauses, binds the plant outputs, and calls the synthesis query~\eqref{eq:query}.

\begin{algorithm}[H]
\caption{Encoding a specification to a synthesis query.}
\label{fig:lowering-algo}
\begin{algorithmic}[1]
\footnotesize
\Function{Encode}{$spec$, plant, workload $w_{comp}$}
  \If{$spec.\texttt{assume}$ declares envelope $\{(v_k, d_k)\}_{k=1}^{n}$}
    \State $\vgp(t) \gets \min\{\, v_k : d_k \ge t \,\}$ \Comment{Eq.~\eqref{eq:wc-trace}}
  \Else
    \State $\vgp \gets spec.\texttt{assume}$ \Comment{already a piecewise lower bound}
  \EndIf
  \State $t_f \gets \min\{\, t : \vgp(t) < 0.9 \,\}$
  \State $t_r \gets \min\{\, t > t_f : \vgp(t) \ge 0.9 \,\}$ \Comment{event times from $\vgp$}
  \State $w \gets (\vgp,\, w_{comp})$ \Comment{disturbance inputs}
  \State $(v_i, p) \gets \text{plant}(u, w)$ \Comment{the plant model mapping $y = g(u,w)$}
  \Statex
  \State $vals \gets \emptyset$
  \ForAll{binding $b \in spec.\texttt{var}$} \Comment{instantiate binding}
    \State $vals \gets vals \cup \{\, b[\,t_f, t_r, w\,] \,\}$ \Comment{e.g.\ $p_{pre} = w_{comp}(t_f)$}
  \EndFor
  \State $\varphi \gets \emptyset$
  \ForAll{clause $c \in spec.\texttt{guarantee}$} \Comment{instantiate clause}
    \State $\varphi \gets \varphi \cup \{\, c[\,t_f, t_r, vals\,] \,\}$
  \EndFor
  \Statex
  
  \State $K \gets \{\, u : P^2 + Q^2 \le S^2 \,\}$ \Comment{apparent-power cap, $S$ from plant}
  \State \Return \Call{Synthesize}{$(v_i,\,p),\,\varphi,\,K,\,w$} \Comment{solve~\eqref{eq:query}}
\EndFunction
\end{algorithmic}
\end{algorithm}

\end{document}